# The polymer mat: Arrested rebound of a compressed polymer layer

Uri Raviv[†], Jacob Klein[†,‡] and T. A. Witten[*]

[†] Department of Materials and Interfaces, Weizmann Institute of Science, Rehovot 76100, Israel.
[‡] Physical and Theoretical Chemistry Laboratory, South Parks Road, Oxford OX1 3QZ, U.K.
[*] James Franck Institute, University of Chicago, Chicago IL 60637 USA

ABSTRACT

Compression of an adsorbed polymer layer distorts its relaxed structure. Surface force measurements from different laboratories show that the return to this relaxed structure after the compression is released can be slowed to the scale of tens of minutes and that the recovery time grows rapidly with molecular weight. We argue that the arrested state of the free layer before relaxation can be described as a Guiselin brush structure[1], in which the surface excess lies at heights of the order of the layer thickness, unlike an adsorbed layer. This brush structure predicts an exponential falloff of the force at large distance with a decay length that varies as the initial compression distance to the 6/5 power. This exponential falloff is consistent with surface force measurements. We propose a relaxation mechanism that accounts for the increase in relaxation time with chain length.

# I. Introduction

Adsorbed polymer layers occur widely in surface phenomena and colloidal dispersions [2-5]. In recent decades it has become possible to predict and to measure the state of polymers in these layers [6-15]. Still, many aspects of these layers have eluded understanding - especially kinetic aspects. One often encounters timescales for equilibration that defy explanation in terms of the characteristic relaxation times of polymers in solution. In this study we examine experiments, in two different laboratories, where these slow relaxations were observed via surface force measurements [16,17]. In both of these cases strong, long range forces are observed upon initial compression to a high volume fraction. But when the surfaces are separated, the forces are much weaker and fall off faster with distance. This compressed force profile persists for many minutes and it persists longer when longer polymers are used. The compressed profile follows an exponential spatial decay of repulsive force over a substantial range of distances.

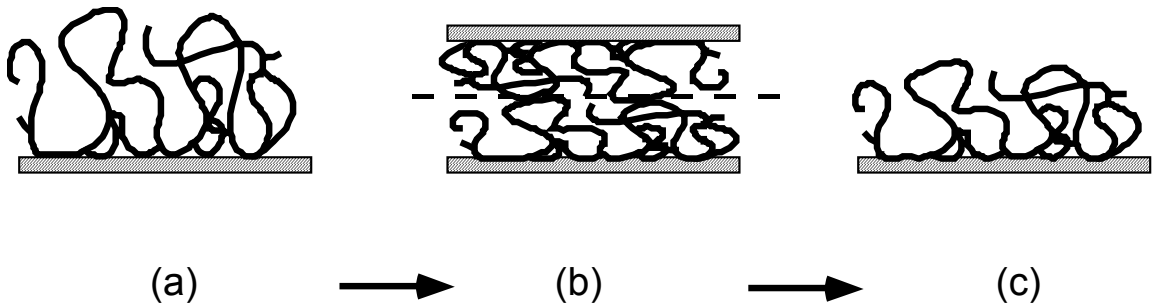

(a) → (b) → (c)

Both of these experiments considered polyethylene oxide (PEO) in a toluene solvent on a mica surface in a range of molecular weights. Qualitatively, similar behavior was observed also with PEO in a 0.1M $KNO_3$ aqueous solution [18]. As the PEO / 0.1M $KNO_3$ solution system is an unusual one [19] and nonetheless similar behavior was observed in both solvents, in a range of molecular weights, we believe that such slow relaxation phenomena occur in a range of adsorbed polymer layers that have been compressed, despite the special character of this polymer. In Section II we describe a form of surface constraint that accounts plausibly for the weakened force in the compressed layer. We then describe the brushlike "mat" structure of the layer when the compression is released. We infer the force profile to be expected from this new structure at short and at long distances. In Section III we compare our predictions with the force measurements in a surface force balance (SFB). The power law predicted at small separations is consistent with the data in that region. At larger distances an exponential falloff is predicted. This agrees well with the experiments, and the observed decay length agrees well with the predicted one arising from the mat structure. We also compare the observed transition distance between weak and strong compression regimes with the predicted thickness of the mat structure and find satisfactory agreement. In the discussion section which follows, we suggest a relaxation mechanism that could account for the increase in relaxation time with molecular weight. We comment on the implications of the mat state for possible applications and suggest further experiments to investigate the mat state more stringently.

## II. Proposed constrained state of a compressed polymer layer

The absorbed polymer layer present before compression is strongly concentrated near each surface [1,12,15,20-22], so that typically, a significant fraction (of order 10-20%) of the surface is covered with monomers. As two surfaces are brought together, the surface concentration increases as further monomers from each polymer layer are pushed onto the surface. In some cases these high concentrations may decrease the monomer mobility markedly [23,24]. Since our experiments show very slow recovery to the initial state, we infer that they suffer this decreased mobility at the surface. Nevertheless, part of the recovery is immediate: the compressed layers are able to expand to several times their compressed size with no observed time lag [16,18], as evidenced by the fact that on separations the measured intersurface forces remain repulsive at all separations $D_{max} > D > 2h_0$, where $2h_0$ is the value of D at closest approach and $D_{max}$ is the value of D when normal steric forces vanish. This behavior leads us to infer that the monomers away from the surface are not immobilized.

These observations lead us to postulate a "mat" state, in which all the monomers in contact with the surface in the initial compressed state are obliged to remain there until some slow relaxation process has occurred. However, the remaining monomers are apparently mobile and free to swell with solvent. The mat thus consists of free loops of polymer between successive wall contacts.

To characterize the behavior of the mat, we must first determine the distribution of loop lengths. We shall take the initial state to be an incompressible melt [25] so that the polymers are Gaussian random walks with step length *a*. The distribution of segment lengths n between successive contacts with a neutral wall is a classic subject of random walk theory [1,26]. The probability p(n) that a loop has length n is proportional [1] to $n^{-3/2}$. This probability distribution is altered by the compressing wall. We take this compressing wall to be the midplane of the gap in the SFB (Sketch 1). This neutral wall simply confines the polymer layer without adsorbing it. This neutral wall increases the probability of contact with the adsorbing wall. For very large n, the probability p(n) must fall off exponentially with n. Since increased n does not bring the chain further from the adsorbing wall, each increment of n brings a constant risk of touching this wall. This exponential falloff evidently sets in when the size of the chain, $an^{1/2}$, reaches the layer confining thickness $h_0$. Thus $p(n) \sim \exp(-n/n_0)$, where $n_0 \approx (h_0/a)^2$ and $a$ is the monomer size. This exponential falloff continues until the loop length n reaches the length of the chain N. From now on we shall assume that N is large enough to be considered infinite.

Such a mat of loops is free to swell substantially in solvent, as first remarked by Guiselin [1]. Guiselin found that this swelling results in a brush-like structure in which the volume fraction $\phi(z)$ at height z falls off as $(a/z)^{2/5}$ in a good solvent. This structure is unlike the original adsorbed layer, in that the surface excess now increases with the maximum length of the loops. In our mat, this maximum is $n_0$, since loops longer than this are exponentially rare. Following Guiselin's treatment [1], we infer a brush thickness $h \approx a n_0^{5/6}$ or $h \approx a (h_0/a)^{5/3}$.

Upon decompression, this mat must exert a restoring force. The force exerted by general Guiselin brushes was worked out by Aubouy et al [22]. Their first step is to determine the volume fraction profile for a surface compressed from the brush thickness h to a smaller thickness D/2. As in a conventional monodisperse brush [27], this $\phi(z)$ retains its unperturbed form for small distances but then becomes constant out to z = D/2. At strong compressions, the unperturbed region becomes negligible so that $\phi(z)$ becomes uniform: $\phi(z) \approx 2\Gamma/\rho D$, where $\Gamma$ is the polymer adsorbance on each surface and $\rho$ is the polymer bulk density (1.13gr/cm$^3$ for PEO). In practice – see Table 1 – the distance of closest approach $h_0$ is, for all polymers studied, related to the adsorbance by a numerical factor, as $\Gamma/\rho \approx (0.5\pm0.1)h_0$. The corresponding osmotic pressure [12,28,29] is $\Pi(D) = kTa^{-3}$ $[\phi(D)]^{9/4} \approx kTa^{-3} (2\Gamma/\rho D)^{9/4}$. From this $\Pi(D)$, we may find the work of compression, $\int dD\ \Pi(D) \approx kTa^{-3} (2\Gamma/\rho)^{9/4} D^{-5/4}$. It is this work that the SFB measures.

The SFB is sensitive enough to detect forces at separations D beyond the high-density thickness regime 2h of the two brushes. Thus, it readily detects surface forces of 20μN/m, which implies a work of compression of kT per (36 nm)$^2$ area. Clearly in this force regime, any polymers producing the force (typically, a few large loops) are expected to be in a dilute state. In this weak force regime, only the longest loops in the mat contribute, and these only contribute when they are longer than usual. In order to find the expected force, we begin by asking how the probability of a loop extending to a height z falls off with z. This probability can be expressed as $p(z) = \Sigma_n p(n)p_n(z)$. The second factor $p_n(z)$, which is the probability that a loop of n monomers expands to height z, is, up to prefactors, equal to $\exp(-U(z)/kT)$, where $U(z)$ is the work required to extend the n-mer loop to a height z. In a good solvent [12], this work may be written $U(z) \approx kT[z/(an^\nu)]^{1/(1-\nu)}$, where $\nu$, the Flory swelling exponent, is roughly 3/5 in a good solvent. As we have seen, the first factor p(n) has the form $p(n) \sim \exp(-n/n_0)$, so that $p(z) \approx \exp[(-n/n_0)-(z/an^\nu)^{1/1-\nu}]$.

From this information, we may find the limiting form of p(z) for large z; in this limit, the sum is dominated by those values of n that minimize this exponent: $n \approx (z/a) n_0^{1-\nu}$. Then $p(z) \approx \exp(-z/D_0)$, where $D_0 \approx a\ n_0^\nu \approx a\ (h_0/a)^{2\nu}$. In the case of a good solvent, $D_0 \approx a (h_0/a)^{6/5}$. Remarkably, the z dependence is a simple exponential, not a stretched exponential.

The osmotic pressure exerted by these dilute chains is proportional to their number. The work of compression is of the same order: it also falls off exponentially with z. We infer that the force seen in the SFB also falls off exponentially with separation D. We summarize the predicted force profile in Sketch 2.

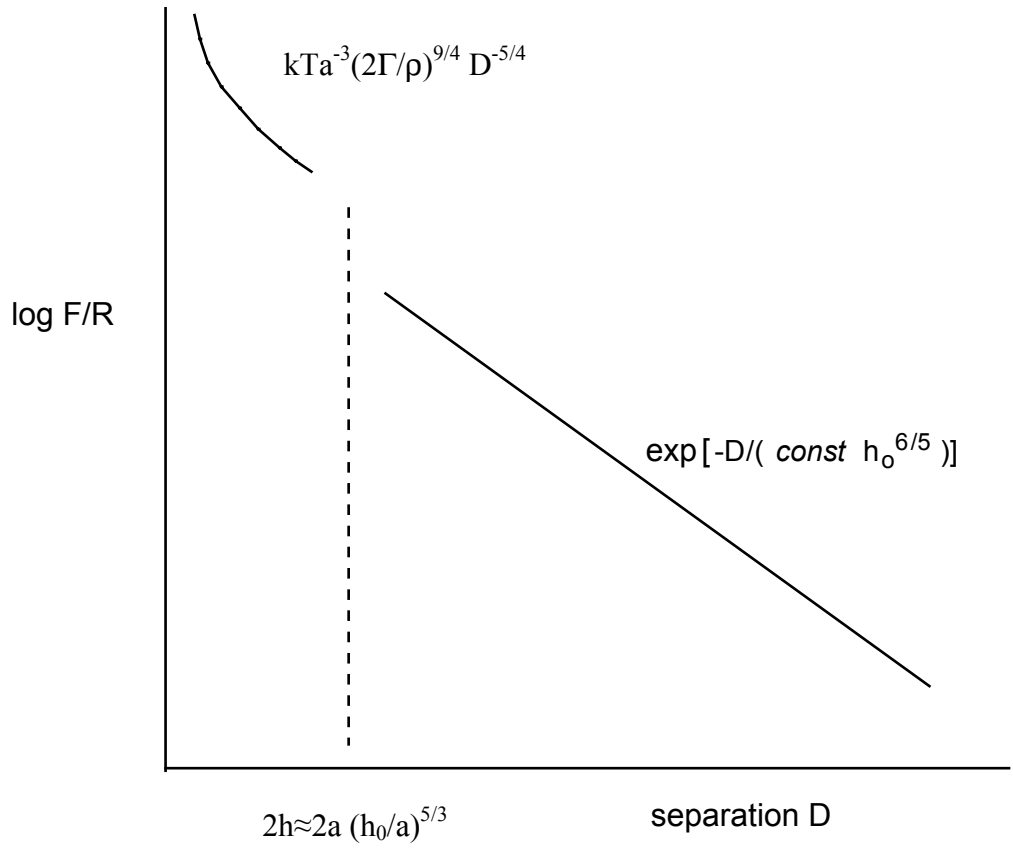

## III. Comparison with measurements

The normal forces $F_n$ between layers of poly(ethylene oxide) (PEO), adsorbed onto smooth, curved solid (mica) surfaces across the good solvent toluene have been determined, in two different laboratories [16,17], as a function of the surface separation D, using a surface force balance (SFB). The SFB measurements were performed with five different molecular weights. Results measured on both compression and decompression are shown, where the force axis is normalized by the radius of curvature R of the mica surfaces, $F_n(D)/R$: in the Derjaguin approximation [30] (for R>>D) $F_n(D)/2\pi R$ is the corresponding interaction energy $E(D)$ per unit area between two flat parallel surfaces, a distance D apart obeying the same force-distance law. This normalization enables comparison of $F_n(D)/R$ profiles from different experiments. In Figure 1-5 the experimental force – distance curves of 5 different molecular weights are shown. The main features are as follows: on initial compression (indicated with solid symbols in the figures) a monotonic repulsion becomes detectable at a range of several $R_g$ (7-9 $R_g$): This repulsion falls off roughly exponentially, with a decay length of $(1.0\pm0.1)R_g$, for all polymers studied. This large decay length indicates weak, marginal adsorption [12,31]. On decompression immediately following close approach (indicated with open symbols in the figures), the forces are considerably shorter-ranged, indicating the forced adsorption of more segments onto the mica surface and a transient compressive distortion of the adsorbed layers. On subsequent recompression the layers relax back to their original (equilibrium) structure to an extent that depends on the molecular weight and the time before re-compression: for the 37 and 40 kg/mole polymers, full relaxation occurs even for 'immediate' recompression (within 10 minutes following the first approach run). For the 112 and 160 kg/mole polymers, the relaxation on immediate re-compression is partial (indicated with cross symbols in figures 3 and 4), and full relaxation to the equilibrium $F_n(D)$ profile occurs only after about one hour. For the 310 kg/mole polymer on immediate re-compression no relaxation can be detected, and full relaxation is obtained only after two hours or more. In figure 1-5 the forces measured on decompression are compared with the long-ranged exponential falloff: $B \exp(-D/D_0)$, where $D_0$ is the decay length, and with the short-ranged power law: $A\, D^{-5/4}$, as predicted by our model (solid lines) (A and B are prefactors). It is impossible to fit the data only to one functional form, as can be realized from the broken lines at each figure. The main feature of the model - long-ranged exponential dependence, is well supported by the data. In particular, the decay lengths predicted are in good agreement with the measurement (see table I). The power law dependence fits the data in most cases, however due to the scatter in the data, it is not the only functional form that could fit the data.

More detailed comparison between the experiments and the model is summarized in table I. The values of $2h_0$ - the closest approach distance, and $2h$ - the distance at which the force appears to go from a power law dependence to an exponential one (sketch 2) are obtained from the experimental data for each molecular weight (figures 1-5). The values of h, which necessarily have a large scatter, are compared with the prediction of our model: $h \approx a(h_0/a)^{5/3}$ ( where $a$ = 0.86±0.14 nm [28]). We obtain that $h/[a(h_0/a)^{5/3}]$ is 1.0±0.3. Slight changes between the conditions of the two sets of experiments (the 37K and 112K [17] and the 40, 160 and 310K [16]) contributed to the scatter in the prefactors. For example the amount of polymer adsorbed on the surfaces was smaller by 10% in the case of 37K compared to the 40K. The exponential decay lengths $D_0$ from figures 1-5 are compared with the values $a^{-1/5}(h_0)^{6/5}$ expected from the model, and we find that $D_0/a^{-1/5}(h_0)^{6/5} = 2.3\pm0.2$, a rather satisfactory fit.

| $M_w$ [kg/mole] | $2\Gamma/\rho$ [nm] | $2h_0$ [nm] ±0.3 | $\Gamma/\rho h_0$ | $2h$ [nm] | $h/[a(h_0/a)^{5/3}]$ | $D_0$ [nm] | $D_0/a^{-1/5}(h_0)^{6/5}$ [nm] |
|---|---|---|---|---|---|---|---|
| 37  | 1.6±0.4[a] | 2.6 | 0.62 | 5±1  | 1.5±0.3 | 2.9±0.2  | 2.1±0.1 |
| 40  | 1.8±0.4[b] | 4.0 | 0.45 | 7±1  | 1.0±0.2 | 6±0.3    | 2.5±0.1 |
| 112 | 2.3±0.5[a] | 4.6 | 0.5  | 11±4 | 1.2±0.4 | 6.8±0.4  | 2.4±0.2 |
| 160 | 2.7±0.6[b] | 8.0 | 0.34 | 17±3 | 0.8±0.3 | 11.5±0.5 | 2.1±0.1 |
| 310 | 3.6±0.8[b] | 8.2 | 0.44 | 17±4 | 0.7±0.4 | 13±0.5   | 2.3±0.1 |

[a] Based on data taken from reference [17].

[b] Based on data taken from reference [16].

## IV. Discussion

The chief support for our mat model for the compressed layer lies in the exponential falloff with distance seen for weak forces. In our model this dependence arises from the exponential distribution of loop lengths. The observed decay length of this exponential is consistent with the expected behavior of the mat structure. Despite this success, some limitations of our conclusions should be noted. First, the scatter in the available data clearly prevents a stringent test of our model; other accounts of the data might prove equally satisfactory. Second, our simplified description has neglected some effects that might be significant. Our assumption of melt conditions in the initial compression is only approximately satisfied, and may well be more valid for some data than for others. Our treatment of the compression of two surfaces as equivalent to confinement by a neutral wall is also approximate. It neglects possible encounters of a chain with the opposite wall and it neglects interpenetration of chains. Still this simple assumption seems adequate for the primitive inferences we have made.

The main puzzle raised by these experiments is the long time scale of the relaxation. This slow relaxation cannot be accounted for fully by locally hindered mobility, since the relaxation becomes slower with increasing chain length N. This N dependence suggests that the relaxation requires co-operative motion of large segments of the chains. We are led to suggest a possible mechanism of this type. In the expanded mat state the loops are stretched. Tension pulls each loop at its anchor points on the surface. Any nonzero mobility at such an anchor point will allow the loop to slip along or through the constraint, thus lengthening the loop. At the final stages of relaxation the layer must resemble an equilibrium adsorbed layer, with a finite fraction of the chain very near the surface. Thus pulling out a loop entails a friction force proportional to the chain length. The expected tension in the largest loops at this stage is of order $kT/(a\,N^\nu)$. The loop is lengthened at a speed v which is the tension divided by the friction factor: $v \sim kT/(a\,N^{1+\nu})$. Finally the time $\tau$ required to equilibrate the loop is of order $N/v \sim N^{2+\nu}$. It remains to be seen whether such strong N dependence will be observed in practice. The progression to this final relaxation appears interesting and rich, as it involves the lengthening of loops in competition with others in an environment of mutual compression.

The mat state suggested here has potentially attractive properties for a solid-liquid interface. Its brush-like structure resembles a grafted polymer layer, though only uniform homopolymers are needed to create it. Such grafted layers often show superior performance for steric stabilization. Recent work [32-34] suggests that grafted layers are particularly effective at supporting large normal loads with low friction. The same virtues might be expected for the original Guiselin brush. But the mat state appears to achieve some of these same virtues without requiring a bulk polymer melt. Indeed our high molecular weight mats showed [17] low friction effects reminiscent of our previous observation with polymer brushes [32].

# Acknowledgement

We thank Albert Johner and Paul F. Luckham for useful discussions and the Eshkol foundation for studentship to U.R. This research was supported by a grant from the United States—Israel Binational Science Foundation (BSF), Jerusalem, Israel and by US National Science Foundation its MRSEC Program under Award Number DMR-980859.

*Tech.*,(2002) submitted). The shear behavior described in ref. 17 also suggests that the adsorption is weak.

Figure Captions

Sketch 1. A) initial adsorbed layer, B) layer compressed to melt density. Midplane, shown as a dashed line is taken as the compressing wall, C) swollen mat state after compressing wall is removed.

Sketch 2. Schematic representation of predicted surface force vs. separation D.

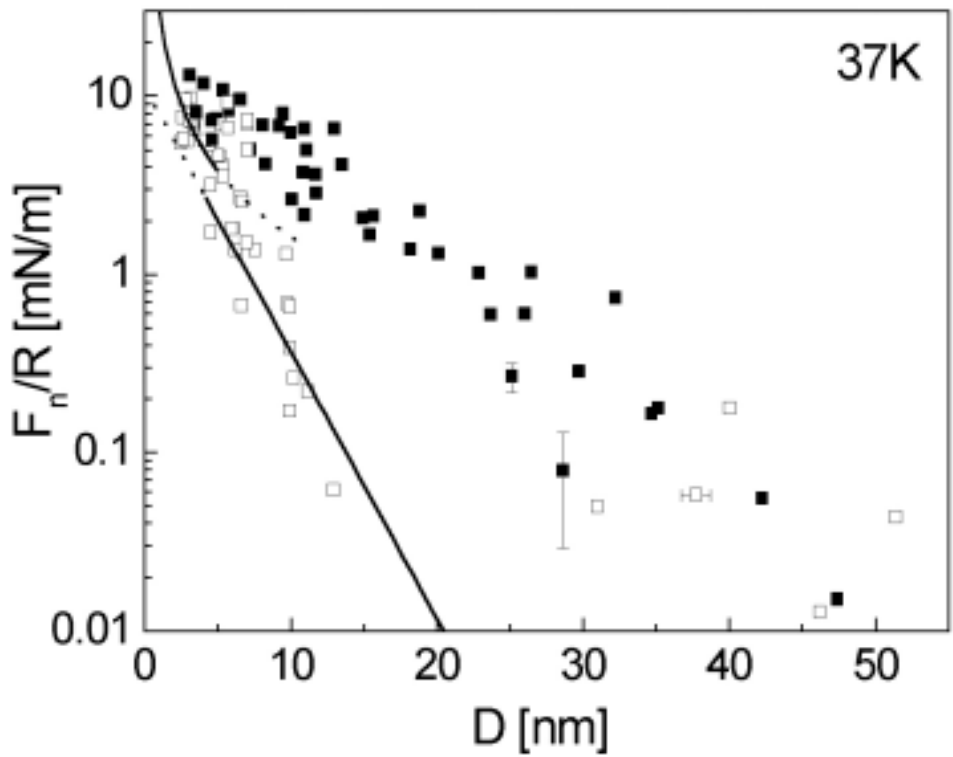

Figure 1

Normal force ($F_n$)-distance (D) profiles between curved mica surfaces in following overnight incubation of the mica surfaces in 100μg/mL solution of PEO ($M_w$ = 37K) in toluene pure toluene, where the force axis is normalized as ($F_n/R$) (R - mean radius of curvature of the mica) to yield the interaction energy per unit area between flat parallel plates obeying the same $F_n(D)$ law, in the Derjaguin approximation[30]. Measurements during compression and rapid decompression of the two surfaces are shown in different sets of experiments. Solid symbols indicate forces measured during compression and open symbols indicate forces measured during decompression. We note that the profiles on a recompression immediately following a decompression are identical – within the scatter – to the original compression profile. The data is taken from reference [17]. The solid lines are plotted where the prediction of the model fits the data well: the power low (in the short range): $A\,D^{-5/4}$, with A = 511.5 mN/m and the exponential decay (in the long-range): $B\exp(-D/D_0)$, with B = 11.5 mN/m and $D_0$ = 2.9 nm. The broken lines are the continuation of the solid lines, in the range at which the same functional form doesn't fit well to the data.

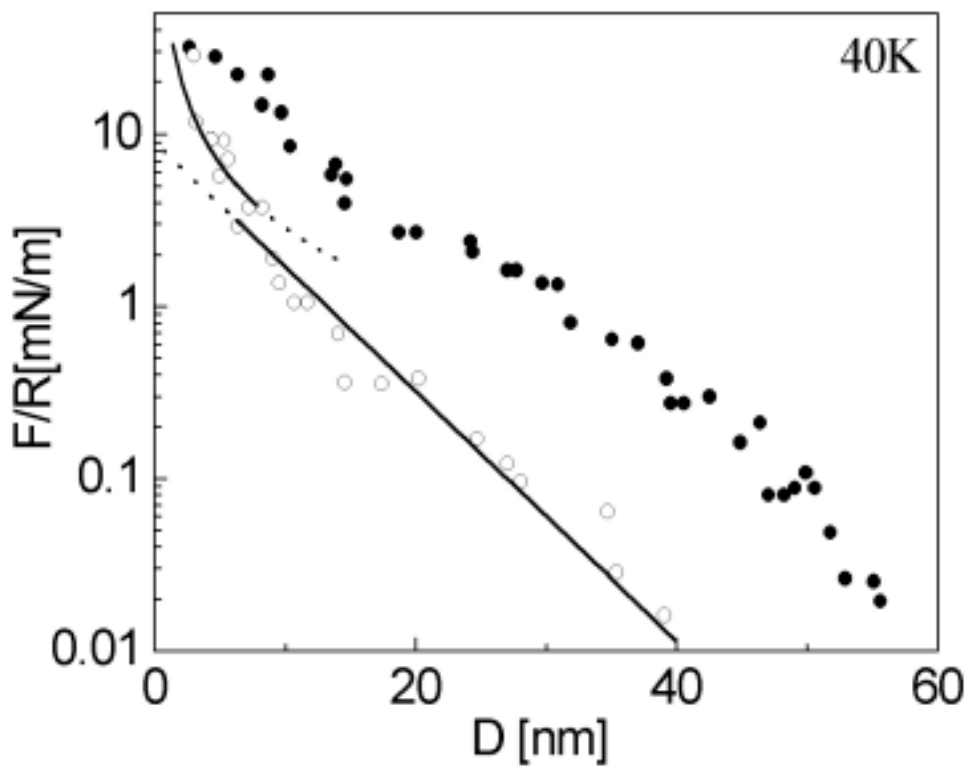

Figure 2.

Normalized force-distance profiles ($F_n/R$) vs. D following overnight incubation of the mica surfaces in 100μg/mL solution of PEO $M_w$ = 40K in toluene, the data is taken from reference [16]. Symbols and lines are as in figure 1, here A = 0.9 N/m, B= 9 mN/m and $D_0$ = 6 nm.

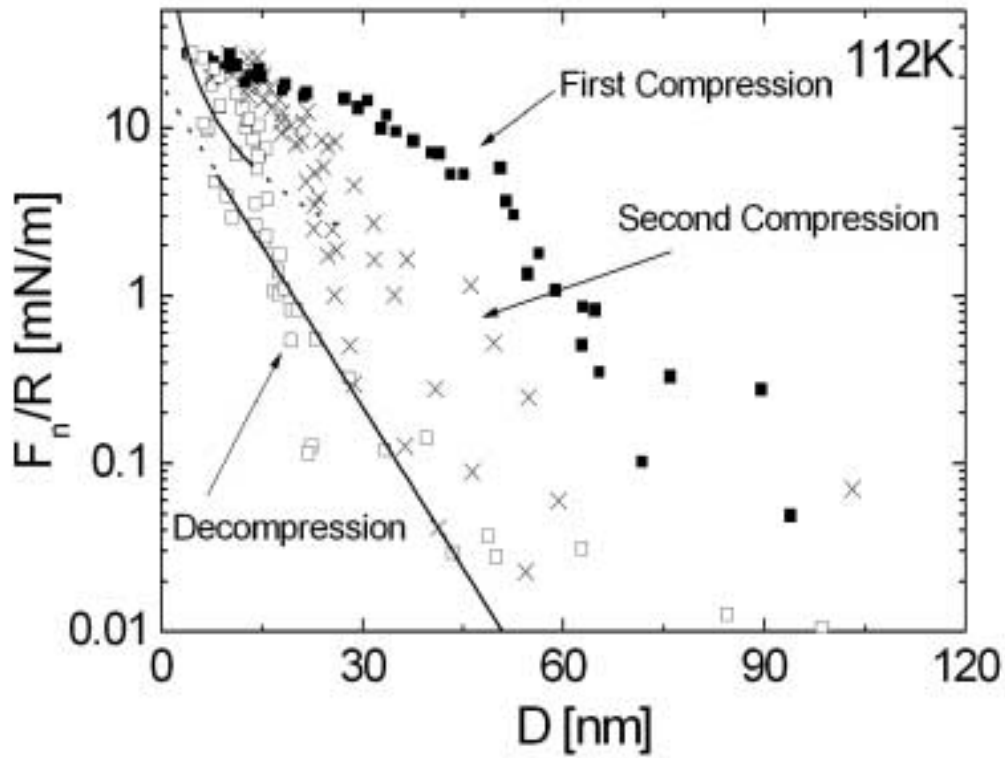

Figure 3.

$F_n/R$ vs. D following overnight incubation of the mica surfaces in 100μg/mL solution of PEO $M_w$ = 112K in toluene, the data is taken from reference [17]. Symbols and lines are as in figure 1, here A = 2.8 N/m, B= 18 mN/m and $D_0$ = 6.8 nm. The cross-like symbols represent forces measured during second compression.

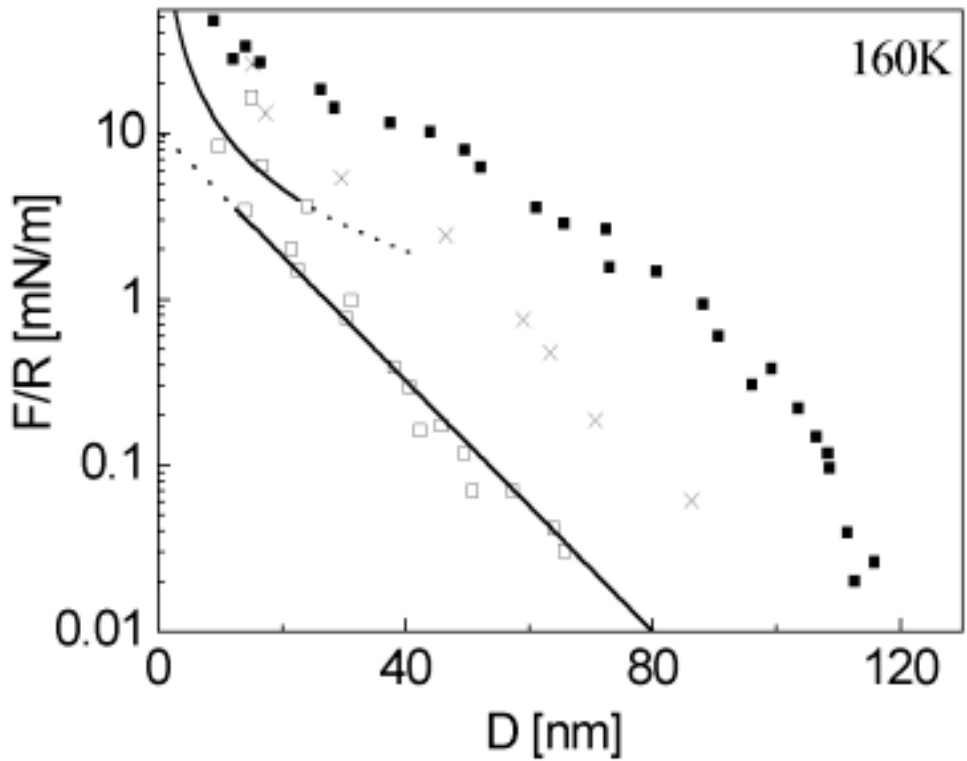

Figure 4.

$F_n/R$ vs. D following overnight incubation of the mica surfaces in 100μg/mL solution of PEO $M_w$ = 160K in toluene, the data is taken from reference [16]. Symbols and lines are as in figure 1, here A = 3.5 N/m, B= 10.5 mN/m and $D_0$ = 11.5 nm. The cross-like symbols represent forces measured during second compression.

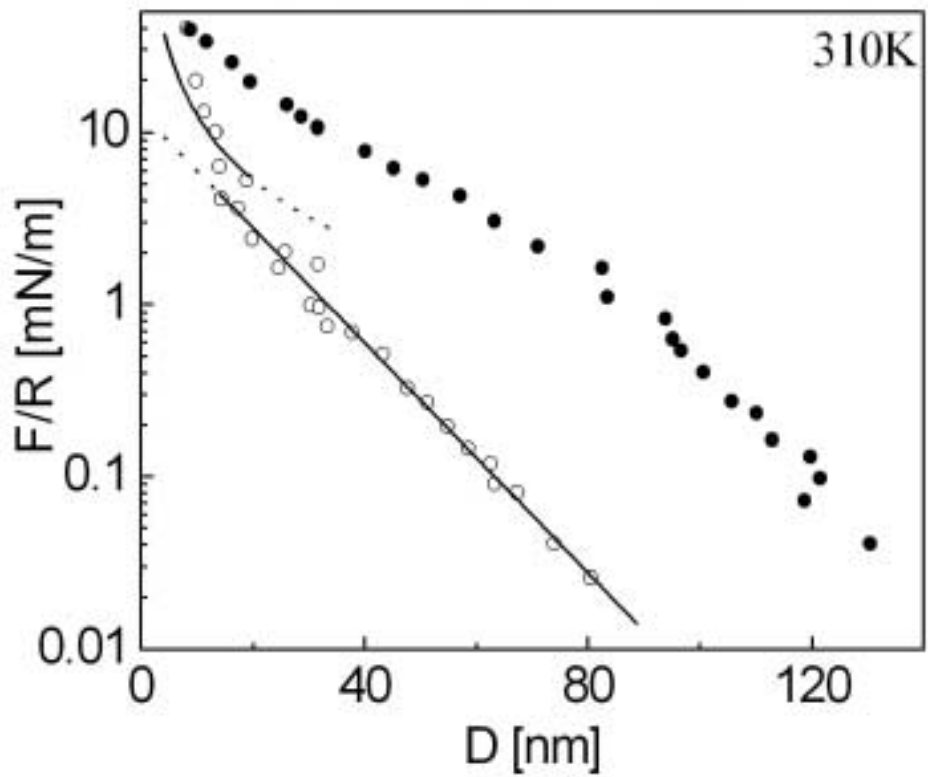

Figure 5.

$F_n/R$ vs. D following overnight incubation of the mica surfaces in 100µg/mL solution of PEO $M_w$ = 310K in toluene, the data is taken from reference [16]. Symbols and lines are as in figure 1, here A = 4 N/m, B= 13 mN/m and $D_0$ = 13 nm. We note that the profiles on a recompression immediately following a decompression are identical – within the scatter – to the decompression profile.